\documentclass[aps,pra,reprint,superscriptaddress,showpacs,floatfix]{revtex4-1}

\usepackage{amsmath}
\usepackage{amssymb}
\usepackage{graphics}
\usepackage{graphicx}
\usepackage{epstopdf}
\usepackage{dcolumn}
\usepackage{bm}
\usepackage{color}

\begin{document}

\title{Partial coherence control delivers skyrmionic topological resilience and transitions}

\author{Yonglei Liu}
\affiliation{Shandong Provincial Key Laboratory of Light Field Manipulation Physics and Applications \& School of Physics and Optoelectronics, Shandong Normal University, Jinan 250358, China}
\affiliation{School of Physical Science and Technology \& Collaborative Innovation Center of Suzhou Nano Science and Technology, Soochow University, Suzhou 215006, China}

\author{Shiqi Chen}
\affiliation{Centre for Disruptive Photonic Technologies, School of Physical and Mathematical Sciences, Nanyang Technological University, Singapore 637371, Singapore}

\author{Zhenyu Guo}
\affiliation{Centre for Disruptive Photonic Technologies, School of Physical and Mathematical Sciences, Nanyang Technological University, Singapore 637371, Singapore}

\author{Kaiqi Zhu}
\affiliation{School of Physical Science and Technology \& Collaborative Innovation Center of Suzhou Nano Science and Technology, Soochow University, Suzhou 215006, China}
\affiliation{Suzhou Key Laboratory of Intelligent Photoelectric Perception, Soochow University, Suzhou 215006, China}

\author{Yahong Chen}
\thanks{Corresponding author: yahongchen@suda.edu.cn}
\affiliation{School of Physical Science and Technology \& Collaborative Innovation Center of Suzhou Nano Science and Technology, Soochow University, Suzhou 215006, China}
\affiliation{Suzhou Key Laboratory of Intelligent Photoelectric Perception, Soochow University, Suzhou 215006, China}

\author{Yangjian Cai}
\thanks{Corresponding author: yangjian\_cai@163.com}
\affiliation{Shandong Provincial Key Laboratory of Light Field Manipulation Physics and Applications \& School of Physics and Optoelectronics, Shandong Normal University, Jinan 250358, China}

\author{Yijie Shen}
\thanks{Corresponding author: yijie.shen@ntu.edu.sg}
\affiliation{Centre for Disruptive Photonic Technologies, School of Physical and Mathematical Sciences, Nanyang Technological University, Singapore 637371, Singapore}
\affiliation{School of Electrical and Electronic Engineering, Nanyang Technological University, Singapore 639798, Singapore}

\author{Fei Wang}
\thanks{Corresponding author: fwang@suda.edu.cn}
\affiliation{School of Physical Science and Technology \& Collaborative Innovation Center of Suzhou Nano Science and Technology, Soochow University, Suzhou 215006, China}
\affiliation{Suzhou Key Laboratory of Intelligent Photoelectric Perception, Soochow University, Suzhou 215006, China}

\date{\today}

\begin{abstract}
Optical skyrmions have recently unlocked topological quasiparticle textures of light, rising in prominence for next-generation ultra-robust information processing. However, to date, their study has been mainly confined to coherent laser fields. Here we extend skyrmions to more general light sources of partially coherent, stochastic optical fields. We define stochastic optical skyrmions and uncover a hidden regime where spatial coherence acts as a primary determinant of topological stability. While environmental randomness typically degrades fully coherent states, we demonstrate that engineered partial coherence provides a self-healing mechanism that preserves topology under extreme turbulence. Moreover, we show that the coherence structure can be actively tailored to trigger on-demand topological phase transitions, such as skyrmion-to-skyrmionium conversion and skyrmion lattice splitting. These findings redefine the boundaries of topological photonics, paving the way for resilient and high-fidelity information platforms that remain operational in general, non-ideal, real-world environments.
\end{abstract}

\maketitle

\section{Introduction}
Skyrmions are topologically protected quasiparticles first introduced in nuclear physics \cite{Skyrme1961,Skyrme1962}. They have become an important concept in condensed matter physics \cite{Khawaja2001,Leslie2009}, especially in magnetism \cite{Bogdanov2020,GobelB2021}, where their robustness and reconfigurability support emerging information technologies \cite{FertA2017,MishraK2025}. Recently, the skyrmion paradigm has been extended to electromagnetic waves \cite{Tsesses2018,Du2019,Shen2024NP}, including free-space optical beams \cite{Gao2020,Gutierrez2021,Shen2021NC,Shen2025AOP}, enabling photonic realizations of topological textures with growing relevance to structured-light control \cite{Guo2020,Lei2021PRL,Wang2024PRL,He2024,Hakobyan2025,Mata2025PRL,Jia2025}, optical computing \cite{WangA2025}, and high-dimensional information processing \cite{WangJ2025,ChenL2025}.

Analogous to magnetic skyrmions, optical skyrmions have been shown to exhibit a degree of topological protection against external perturbations in harsh environments, including complex media, biological tissues, and turbulent atmospheres \cite{WangA2024, OrnelasP2025, PiresD2026, Peters2025, GuoZ2025, PetersC2026}. However, under strong atmospheric turbulence the skyrmion number, the key topological invariant used to quantify the skyrmionic texture, can degrade and become realization-dependent. Recovering the correct skyrmion number may then require long integration times \cite{Peters2025} or extensive post-processing \cite{GuoZ2025, PetersC2026, Peters2026AP}, which is incompatible with scenarios that demand rapid, real-time readout. This limitation is particularly restrictive for practical applications such as topology-encoded free-space optical communications, where information is carried by the skyrmion number and must be decoded on the fly.

A natural route to overcome this challenge is to move beyond idealized laser fields \cite{Gbur14}. Nearly all optical skyrmions reported to date have been created and studied in fully coherent and fully polarized beams. Yet natural optical fields are intrinsically stochastic: partial coherence and random perturbations are the rule rather than the exception \cite{Mandel1995,Wolf2007}. Partially coherent beams have been widely studied and have shown distinct advantages in suppressing turbulence-induced fluctuations, improving beam transport, and enhancing the robustness of free-space optical communications \cite{Gbur14, Wang15Propagation}. This raises a fundamental question: are optical skyrmions a generic feature of partially coherent light, and can partial coherence endow them with enhanced resilience to environmental randomness such as atmospheric turbulence?

Here we establish the concept of partially coherent optical skyrmions. We demonstrate stochastic skyrmion textures and show that their topology can be preserved during propagation through spatial-coherence engineering. Remarkably, partially coherent skyrmions remain robust even under strong turbulence---without requiring long integration times or extensive post-processing---marking a key step toward real-time, topology-encoded photonic links in complex environments. Moreover, we uncover coherence-driven topological transitions in the low-coherence regime, including skyrmion-to-skyrmionium conversion and the controlled evolution from isolated skyrmions to ordered skyrmion lattices. Together, these results extend optical skyrmions from ideal coherent laser fields to the more universal domain of partially coherent light, opening routes to turbulence-resilient topological photonics for free-space communications, sensing, and photonic information processing.

\section{Results}
\subsection{Concept}
\begin{figure*}[h!]
\centering
\includegraphics[width=0.84\linewidth]{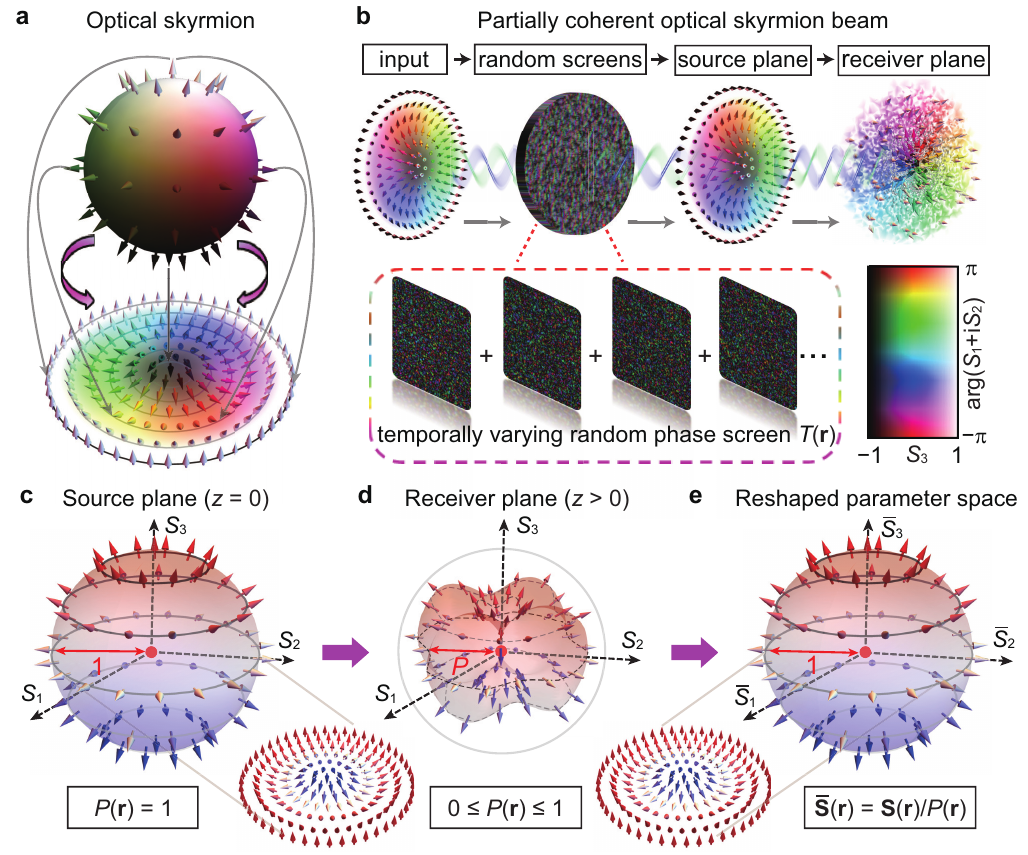}
\caption{\textbf{Concept and generation of partially coherent optical skyrmions}. (a) Optical skyrmions correspond to a topological mapping between the real-space polarization texture and a parameter space represented by the Poincar\'e sphere. (b) A partially coherent skyrmion beam is generated by transmitting a fully coherent (deterministic) skyrmion through temporally varying, polarization-insensitive random phase screen $T(\mathbf{r})$ and performing an ensemble (time) average. (c) At the source plane ($z=0$), the field is fully polarized with $P(\mathbf{r})=1$, so the Stokes vectors lie on the unit Poincar\'e sphere. (d) After propagation ($z>0$), spatial partial coherence leads to inhomogeneous depolarization ($0\le P(\mathbf{r})\le 1$), causing the Stokes-vector locus to shrink into the interior of the unit sphere. (e) A reshaped parameter-space representation is obtained by local renormalization of the Stokes vector, $\overline{\mathbf{S}}(\mathbf{r})=\mathbf{S}(\mathbf{r})/P(\mathbf{r})$, which restores a unit-sphere mapping for characterizing the skyrmion topology.}
\label{figure1}
\end{figure*}

Skyrmion textures in optics arise from a continuous, topological mapping between an internal parameter space and the real-space distribution of an optical field \cite{Shen2024NP}, as schematically illustrated in Fig.~\ref{figure1}(a). Such textures have been constructed using various degrees of freedom of light, including the electric field \cite{Tsesses2018}, spin angular momentum \cite{Du2019}, Poynting vector \cite{Wang2024PRL}, and even scalar evanescent components \cite{Mata2025PRL}. For fully coherent beams, the most widely explored realization is the Stokes skyrmion \cite{Gao2020}, in which the parameter space is the Poincar\'e sphere spanned by the (normalized) Stokes vector $\mathbf{S}(\mathbf{r})$, and the corresponding real-space texture is encoded in the spatially varying polarization state across the transverse beam profile. A canonical platform is provided by full Poincar\'e beams \cite{Beckley2010}, whose transverse field can be constructed as a coherent superposition of two orthogonally polarized Laguerre--Gaussian (LG) modes with, in general, different radial and azimuthal indices,
\begin{equation}
\mathbf{E}(\mathbf{r}) = a_1 \mathrm{LG}^{l_1}_{p_1}(\mathbf{r})\hat{\mathbf{e}}_1
+ a_2 e^{i\gamma} \mathrm{LG}^{l_2}_{p_2}(\mathbf{r})\hat{\mathbf{e}}_2,
\label{eqE}
\end{equation}
where $\mathbf{r}$ is the two-dimensional coordinate in a transverse plane, perpendicular to the beam propagation axis. $\mathrm{LG}^{l}_{p}$ denotes the complex amplitude of an LG mode with radial index $p$ and azimuthal index $l$, $a_{1,2} \geq 0$ set the relative modal weight (with amplitude ratio $\eta=a_2/a_1$), $\gamma$ is the relative phase, and $\hat{\mathbf{e}}_{1,2}$ are two orthogonal unit polarization basis vectors. When $|l_1|\neq |l_2|$, the field in Eq.~(\ref{eqE}) supports skyrmion features in the Stokes texture $\mathbf{S}(\mathbf{r})$ {(see Methods)}. The topology of a given skyrmion configuration is quantified by the skyrmion number \cite{Shen2024NP}
\begin{equation}
\label{eqNsk}
N_\mathrm{sk} = \dfrac{1}{4\pi} \iint_\sigma \mathbf{S}(\mathbf{r}) \cdot \left[ \dfrac{\partial \mathbf{S}(\mathbf{r})}{\partial x} \times \dfrac{\partial \mathbf{S}(\mathbf{r})}{\partial y} \right] \mathrm{d}^2 \mathbf{r},
\end{equation}
where $\sigma$ denotes the region over which the skyrmion is defined. Physically, $N_\mathrm{sk}$ counts how many times the Stokes vector wraps the unit parameter sphere as $\mathbf{r}$ spans $\sigma$.

To generate a partially coherent skyrmion field, as illustrated in Fig.~1(b), one may conceptually pass a fully coherent skyrmion field through a temporally varying, polarization-insensitive random phase screen. Such a dynamic screen imprints a time-dependent, spatially random phase on the beam, thereby reducing the spatial coherence while preserving the underlying polarization structure \cite{DongZ2022,HuZ2025}. This picture is closely related to free-space propagation through atmospheric turbulence \cite{Andrews1998}, which is often modeled as a sequence of polarization-insensitive random phase screens. For statistically stationary screens, the second-order statistical properties of the resulting partially coherent vector field are described by the cross-spectral density (CSD) matrix \cite{Mandel1995}
\begin{equation}
\label{eqCSD}
\mathbf{W}(\mathbf{r}_{1},\mathbf{r}_{2}) = \mathbf{E}^\ast(\mathbf{r}_1)\mathbf{E}^{\mathrm{T}}(\mathbf{r}_2)\,\mu(\mathbf{r}_1, \mathbf{r}_2),
\end{equation}
where $\mu(\mathbf{r}_1,\mathbf{r}_2)$ is the complex degree of spatial coherence between the points $\mathbf{r}_1$ and $\mathbf{r}_2$, and $\ast$ and $\mathrm{T}$ denote complex conjugation and transpose, respectively. The spatial structure of $\mu(\mathbf{r}_1, \mathbf{r}_2)$ is determined by the correlation of the random screen, i.e., $\mu(\mathbf{r}_1, \mathbf{r}_2)\propto \langle T^\ast(\mathbf{r}_1)T(\mathbf{r}_2)\rangle$, where $\langle\cdot\rangle$ denotes a time average and $T(\mathbf{r})$ is the screen transmission function. In the experiment, however, the partially coherent skyrmion fields are not produced by first creating a deterministic coherent skyrmion and then passively decohering it in a fluctuating medium. Instead, they are synthesized directly at the source plane using the random-mode superposition method. In this approach, each random realization contains the prescribed vector-mode structure, while the ensemble of realizations determines the designed spatial coherence function. Thus, both the skyrmionic topology and the coherence properties are engineered at the source. Experimental details are provided in Methods.

Equation~(\ref{eqCSD}) implies that the partially coherent skyrmion source shares the same ensemble-averaged polarization structure as its fully coherent counterpart in Eq.~(\ref{eqE}). In particular, because the random phase screen is polarization-insensitive, it does not alter the averaged polarization matrix at the source plane, and hence yields the same Stokes texture $\mathbf{S}(\mathbf{r})$. As a result, both the fully coherent and the partially coherent fields exhibit the same skyrmion configuration at the source: the Stokes vectors lie on the surface of the unit Poincar\'e sphere, as illustrated in Fig.~1(c).

Upon propagation, however, the presence of spatial partial coherence at the source leads to a reduction of the local degree of polarization at the observation plane \cite{DongZ2022, HuZ2025}. Consequently, the measured Stokes vectors no longer remain on the unit sphere but shrink into its interior, reflecting a {partially polarized} field. Importantly, unlike the case of a spatially uniform (global) mixed state \cite{OrnelasP2025}, the depolarization induced by spatial partial coherence is generally {inhomogeneous}: the degree of polarization $P(\mathbf{r})$ varies across the beam profile. Geometrically, this corresponds to a nonuniform inward ``denting'' of the Stokes-vector locus on the Poincar\'e sphere, with the radial distance to the sphere center given by $P(\mathbf{r})$, as sketched in Fig.~1(d).

To define an appropriate parameter-space representation for such inhomogeneously partially polarized skyrmion fields, we locally renormalize the Stokes vector as
\begin{equation}
\label{normalization1}
\overline{\mathbf{S}}(\mathbf{r}) = \frac{\mathbf{S}(\mathbf{r})}{P(\mathbf{r})},
\end{equation}
so that $|\overline{\mathbf{S}}(\mathbf{r})|=1$ pointwise. This normalization restores a unit-sphere mapping and provides a well-defined parameter sphere for characterizing the skyrmion texture, as shown in Fig.~\ref{figure1}(e). Accordingly, the skyrmion number is evaluated from Eq.~(\ref{eqNsk}) by replacing $\mathbf{S}(\mathbf{r})$ with $\overline{\mathbf{S}}(\mathbf{r})$.

It is important to note that Eq.~(\ref{normalization1}) becomes ill-defined when the local degree of polarization approaches zero, i.e., $P(\mathbf{r}) \to 0$. Physically, this limit corresponds to a locally unpolarized field, for which the polarization state is completely randomized and no well-defined local Stokes vector can be assigned. In other words, as $P(\mathbf{r})$ decreases, the skyrmionic Stokes texture encoded in the partially coherent field is progressively buried in the unpolarized background; in the limiting case $P(\mathbf{r})=0$, the local skyrmionic texture disappears.

In extreme situations, such as very low spatial coherence or very strong turbulence, the degree of polarization may be strongly degraded. In this case, the topological texture may be overwhelmed by the unpolarized background and random noise, and the skyrmion number can no longer be reliably extracted without additional signal recovery. Experimentally, such weak polarization textures could in principle be accessed by improving the polarization signal-to-noise ratio, for example through higher-sensitivity polarimetric detection, longer averaging over source fluctuations, or other polarization-signal recovery techniques. These extreme cases, however, lie outside the parameter regime investigated in the present work.

We further note that the definition in Eq.~(\ref{normalization1}) is generally applicable to partially coherent and partially polarized optical fields, irrespective of their specific spatial coherence structure, provided that the local degree of polarization remains finite in the region where the topology is evaluated.

\subsection{Partially coherent skyrmion beam}
To investigate the propagation dynamics of partially coherent skyrmion beams, we first synthesize a skyrmion field with a Gaussian degree of spatial coherence \cite{Wolf2007}, i.e., $\mu(\mathbf{r}_1,\mathbf{r}_2)=\exp\!\left[-(\mathbf{r}_1-\mathbf{r}_2)^2/(2\delta_0^2)\right]$, where $\delta_0$ controls the transverse coherence width. Experimental details of the synthesis and characterization are provided in {Methods}. The input field is formed by superposing a right-handed $\mathrm{LG}^{1}_{0}$ mode and a left-handed $\mathrm{LG}^{0}_{0}$ mode in two orthogonal polarization channels, with equal modal weights $\eta=1$, see in Fig.~2(a), and with the coherence width of $\delta_0=0.5~\mathrm{mm}$. Figures~2(b)--(e) show the measured Stokes parameters and the corresponding Stokes texture $\overline{\mathbf{S}}(\mathbf{r})$ of a partially coherent skyrmion beam during focusing by a thin lens. The corresponding theoretical results are given in Supplementary Sec.~S3. The focal plane is used to represent the far-field distribution of the beam. At the source plane, the Stokes texture exhibits a N\'eel-type skyrmion, with a measured skyrmion number $N_\mathrm{sk}=-0.8789$. The deviation from the ideal value $N_\mathrm{sk}=-1$ arises from the finite support used in the experimental evaluation: $N_\mathrm{sk}$ is computed over the region where the intensity exceeds $2\%$ of the peak value, whereas the theoretical value corresponds to integration over the entire transverse plane.

Upon propagation from the source plane to the focal plane, the transverse patterns of the (unnormalized) Stokes parameters $s_1$ and $s_2$ undergo a clockwise rotation by approximately $\pi/2$, which transforms the Stokes texture from the N\'eel-type into a Bloch-type configuration. Importantly, the skyrmion number remains effectively unchanged throughout this evolution, consistent with its topological character. By varying the spatial coherence width of the beam, this N\'eel-to-Bloch conversion is observed to follow the same evolution. Although reducing the spatial coherence leads to progressive depolarization during propagation, as evidenced in Fig.~2(g) by the decrease of the polarized power fraction $P_p$ from unity, the skyrmion number $N_\mathrm{sk}$, defined via the renormalized Stokes field, remains effectively stable, as shown in Fig.~2(f). The slight decrease of $N_\mathrm{sk}$ observed during propagation is mainly attributed to the finite intensity region used in the experimental evaluation.

We find that the free-space robustness of partially coherent skyrmion textures depends sensitively on the amplitude ratio $\eta$ between the two orthogonally polarized constituents. For example, at $\eta=1.25$ [Fig.~3(a)--(d)], the skyrmion number drops from $N_\mathrm{sk}=-0.9226$ at the source plane to $N_\mathrm{sk}=0.1296$ at the focal plane, indicating an abrupt loss of the skyrmion texture. This strong $\eta$ dependence can be understood from the distinct propagation behaviors of the partially coherent $\mathrm{LG}^{1}_{0}$ and $\mathrm{LG}^{0}_{0}$ modes in the two polarization channels. Unlike a fully coherent $\mathrm{LG}^{1}_{0}$ beam, which maintains a dark on-axis core, a partially coherent $\mathrm{LG}^{1}_{0}$ beam gradually builds up on-axis intensity during propagation \cite{Wang09partially}. As a result, beyond a critical distance the on-axis intensity of the partially coherent $\mathrm{LG}^{1}_{0}$ component can exceed that of the partially coherent $\mathrm{LG}^{0}_{0}$ component [see Fig.~3(e)]. Once this crossover occurs, the sign of the central $\overline{S}_3$ reverses from $-1$ to $1$ [see Fig.~3(f)], triggering an abrupt breakdown of the skyrmion topology. Consistently, Fig.~3(g) shows that $\overline{S}_3(\mathbf{r})$ no longer spans the full range $[-1,1]$ at the corresponding propagation distance, so the Stokes mapping fails to cover the entire parameter sphere and the topological texture becomes corrupted. Notably, the above physical mechanism also applies to partially coherent skyrmion beams with higher-order azimuthal indices, i.e., when both $l_1$ and $l_2$ are nonzero. A detailed analysis of the critical propagation distance is provided in Supplementary Sec.~S1.

As $\eta$ increases further, the decay of $N_\mathrm{sk}$ during focusing becomes progressively faster, as summarized in Fig.~3(h). By contrast, for $\eta \leq 1$ the skyrmion texture remains intact throughout propagation because the on-axis intensity of the $\mathrm{LG}^{0}_{0}$ component stays larger than that of the $\mathrm{LG}^{1}_{0}$ component [Fig.~3(e)], preventing the central $\overline{S}_3$ inversion [Fig.~3(f)]. Additional theoretical results for different $\eta$ are given in Supplementary Fig.~1. Further experimental and theoretical propagation dynamics for other partially coherent skyrmion textures are presented in Supplementary Figs.~2--4, confirming the stability of partially coherent skyrmion beams in free space under the condition $\eta \leq 1$.

\subsection{Robustness of partially coherent optical skyrmions in turbulence}
In recent years, increasing attention has been paid to the stability of skyrmion topological textures when structured beams propagate through harsh environments such as atmospheric turbulence \cite{WangA2024,PiresD2026,Peters2025,GuoZ2025,PetersC2026}. For a fully coherent skyrmion beam, one may naively expect the skyrmion number to remain invariant under turbulence, since turbulence is often modeled as a continuous phase perturbation. Here we show that this expectation can fail in the strong-turbulence regime, where diffraction accumulated along the turbulent channel cannot be neglected \cite{Andrews1998}. In this case, turbulence-induced diffraction can generate additional singular structures, leading to pronounced degradation and realization-to-realization instability of the skyrmion number. We note that, under strong turbulence, the skyrmion number can be partially recovered by ensemble averaging over many turbulence realizations \cite{Peters2025} or by extensive post-processing of the measured data \cite{GuoZ2025,PetersC2026, Peters2026AP}. However, such approaches are time-consuming and therefore impractical for applications requiring rapid, real-time decoding of topology in complex environments.

In what follows, we demonstrate that partially coherent skyrmion beams can deliver a stable realization-to-realization skyrmion-number output even under strong atmospheric turbulence. The key point is that, in the partially coherent case, the detector records an ensemble-averaged Stokes texture over many source realizations within a single quasi-static turbulence realization. This corresponds to the time-scale condition $\tau_s\ll\tau_d\ll\tau_t$, where $\tau_s$ is the fluctuation time of the partially coherent source, $\tau_d$ is the detector integration time, and $\tau_t$ is the characteristic evolution time of the turbulence. Under this condition, the detector averages over the fast source fluctuations but not over different turbulence realizations. Thus, a single set of Stokes measurements within one turbulence realization is sufficient to obtain a stable skyrmion number. This single-realization stability is the practical advantage of partial coherence in our work.

This distinction highlights the central message of our work: whereas previous studies have mainly explored the robustness of optical topology to randomness \cite{Peters2025}, here we show that engineered randomness, in the form of controlled partial coherence, can itself serve as a resource for stabilizing optical skyrmion topology. In this sense, our work addresses not only ``robustness to randomness'', but also ``robustness via randomness''.

We first simulate a fully coherent skyrmion beam synthesized from two orthogonally polarized $\mathrm{LG}^{0}_{0}$ and $\mathrm{LG}^{2}_{0}$ modes propagating through a $1000~\mathrm{mm}$ turbulent channel. The turbulence is modeled with a multi-phase-screen approach \cite{Schmidt2010, Peters2025AOP} that explicitly includes diffraction between successive screens (see Supplementary Sec.~S4). At the receiver plane we evaluate the Stokes parameters, the skyrmion number, and the complex Stokes field $\overline{S}_\mathrm{c}=\overline{S}_1+i\overline{S}_2$ \cite{Peters2025} together with the corresponding mapping of $\overline{\mathbf S}(\mathbf r)$ on the reshaped parameter sphere. The phase singularities in the complex Stokes field, which occur at points where $\overline{S}_1=\overline{S}_2=0$ and hence $\overline{S}_3=\pm1$, correspond to circular-polarization singularities, i.e., C-points. These singularities are directly connected to the skyrmion number through Stokes' theorem \cite{McWilliamA2023}. Although the phase modulation on each screen is continuous, Figs.~4(a)--(c) show that diffraction produces additional phase singularities in $\arg(\overline{S}_c)$ at the output plane as the turbulence strength increases (parameterized by the refractive-index structure constant $C_n^2$). These extra singularities are accompanied by an increasingly irregular and fragmented mapping of $\overline{\mathbf S}(\mathbf {r})$ on the parameter sphere, and correlate with a clear degradation of the skyrmion number. Under strong turbulence, e.g., $C_n^2=1\times10^{-8}~\mathrm{m^{-2/3}}$, different turbulence realizations yield substantial shot-to-shot fluctuations: as shown in Fig.~5(a), the output skyrmion number becomes unstable with mean $\overline{N}_\mathrm{sk}=-1.6497$ and standard deviation $\sigma=0.6431$. We note that the parameter-sphere coverage plots and the phase-singularity maps of $\overline{S}_\mathrm{c}$ provide complementary views of the same topological degradation process. The former visualizes the global mapping from real space to the parameter sphere, whereas the latter reveals the underlying singular skeleton that governs the phase winding and contributes to the skyrmion number.

Strikingly, reducing the spatial coherence of the incident beam markedly enhances the topological robustness in turbulence. Although the Stokes-parameter patterns are still distorted by turbulence (see Supplementary Fig.~6), both the proliferation of phase singularities in $\overline{S}_\mathrm{c}$ and the irregular mapping on the parameter sphere are progressively suppressed when the coherence width becomes comparable to or smaller than the beam waist, as shown in Figs.~4(d)--(f) and 4(g)--(i). This suppression mitigates the turbulence-induced degradation of $N_\mathrm{sk}$. The dependence of the stability of $N_\mathrm{sk}$ on both the spatial coherence width and the beam waist is further demonstrated in Supplementary Sec.~S5. Consistently, Figs.~5(b) and 5(c) show that partially coherent skyrmion beams exhibit a highly stable skyrmion number across different turbulence realizations. We note that $C_n^2=1\times10^{-8}~\mathrm{m^{-2/3}}$ corresponds to the strong-fluctuation regime, as indicated by the intensity scintillation index (SI) of the fully coherent case [$\mathrm{SI} =1.3223>1$; Fig.~5(a)]. Although SI decreases with reduced coherence [Figs.~5(b) and (c)], it remains above unity even for $\delta_0=0.5~\mathrm{mm}$, confirming that the observed stabilization of $N_\mathrm{sk}$ is not a consequence of entering a moderate or weak-turbulence regime.

The underlying mechanism can be understood as follows. A partially coherent beam can be viewed as an incoherent superposition of many spatial modes (or, equivalently, a statistical ensemble of fields) within a single turbulence realization \cite{Shirai03, Gbur14}. This modal incoherent averaging suppresses the formation and survival of singularities in $\overline{S}_c$ and smooths the parameter-sphere mapping of $\overline{\mathbf S}(\mathbf r)$, thereby stabilizing the skyrmion number. In principle, a similar averaging could be achieved by time-averaging over many independent turbulence realizations \cite{Peters2025}, but such an approach would require long acquisition times and is therefore impractical for applications that rely on rapid readout of skyrmion topology. In contrast, partial coherence can be generated and modulated rapidly, enabling real-time production of turbulence-resilient skyrmion states of light.

Finally, we emphasize that this coherence-enabled robustness of the skyrmion number is distinct from the behavior of the total orbital angular momentum (OAM). In our simulations, the total OAM exhibits pronounced turbulence-induced fluctuations regardless of the input spatial coherence [Figs.~5(d)--(f)], in clear contrast to the stabilized skyrmion topology observed for partially coherent beams.

\subsection{Topological transitions induced by optical coherence}
Beyond its advantage in mitigating turbulence-induced degradation, optical coherence provides an additional degree of freedom to actively reshape skyrmion textures and even induce topological transitions during propagation. We demonstrate this capability with two classes of engineered non-Gaussian coherence functions.

We first consider a partially coherent skyrmion beam whose degree of spatial coherence takes a Laguerre--Gaussian (LG) correlation form \cite{Yu2023},
\begin{equation}
\mu(\mathbf{r}_1,\mathbf{r}_2)=
L_n^{0}\!\left[-\frac{(\mathbf{r}_1-\mathbf{r}_2)^2}{2\delta_0^2}\right]
\exp\!\left[-\frac{(\mathbf{r}_1-\mathbf{r}_2)^2}{2\delta_0^2}\right],
\end{equation}
where $L_n^{0}$ is the generalized Laguerre polynomial of radial order $n$ (zero azimuthal order) and $\delta_0$ sets the coherence width [see the spatial distribution in Fig.~6(a)]. Under this distinctive coherence structure, the Stokes-vector texture exhibits a pronounced propagation-induced topological transition [Figs.~6(b)--(e)]: starting from a N\'eel-type skyrmion at the source plane ($z=0$), the texture evolves into a skyrmionium at an intermediate distance around $z=0.9f$, where the skyrmion number approaches $N_\mathrm{sk}\approx 0$. Such a skyrmion--skyrmionium conversion is generally absent in fully coherent skyrmion beams under smooth phase perturbations \cite{FanX2025, ZhenW2026}. The propagation distance at which the transformed topological texture becomes fully developed depends on the engineered spatial coherence. In general, lower spatial coherence accelerates the coherence-induced reshaping process, so the transition can occur at a shorter propagation distance.

At the focal plane, the skyrmionium consists of two nested Bloch-type skyrmions with opposite polarity: a central component with polarity $+1$ and an outer component with polarity $-1$. The experimentally evaluated skyrmion number at focus is $N_\mathrm{sk}=0.2884$, which is primarily due to the finite integration domain used in the measurement (restricted to regions with intensity above $2\%$ of the peak value). Extending the integration to the full transverse plane yields a skyrmion number close to zero, consistent with the skyrmionium character.

The physical origin of this extraordinary transition lies in the discontinuous phase structure encoded in the engineered coherence function. As shown in Fig.~6(a), the phase of $\mu(\mathbf{r}_1,\mathbf{r}_2)$ contains a circular $\pi$-phase dislocation, implying that the source experiences an effective phase-discontinuous perturbation in a statistical, ensemble-averaged sense. During propagation, the engineered spatial coherence structure modifies the evolution of the polarization matrix and thereby reshapes the Stokes texture. Consequently, the associated topological transitions do not appear at the source plane, but emerge progressively during propagation. These coherence-induced structures become most clearly developed in the far field, which corresponds to the focal plane in our experiment, as shown in Fig.~6. This behavior is in sharp contrast to a deterministic $\pi$-phase step imprinted directly on a fully coherent field.

As a second example, we engineer a Hermite--Gaussian (HG) degree of coherence \cite{Yu2023},
\begin{equation}
\mu(\mathbf{r}_1,\mathbf{r}_2)=
H_m\!\left[\frac{x_1-x_2}{\sqrt{2}\,\delta_0}\right]\,
H_n\!\left[\frac{y_1-y_2}{\sqrt{2}\,\delta_0}\right]\,
\exp\!\left[-\frac{(\mathbf{r}_1-\mathbf{r}_2)^2}{2\delta_0^2}\right],
\end{equation}
where $H_{m}$ and $H_{n}$ are Hermite polynomials of orders $m$ and $n$, respectively. The amplitude of this coherence function forms a rectangular spot array, while the phase alternates by $\pi$ between neighboring spots [Fig.~6(f)], indicating a coherence phase landscape with rectangular-symmetry discontinuities.

Under this coherence structure, the Stokes texture gradually splits into four identical, spatially separated skyrmion parts during propagation [Figs.~6(g)--(j)]. Correspondingly, the skyrmion number evolves from $N_\mathrm{sk}=-0.8709$ at the source plane to $N_\mathrm{sk}=-3.7818$ in the focal plane (far field), with each sub-beam contributing approximately $-0.94$. This topological multiplication originates from the rectangularly symmetric, phase-discontinuous coherence landscape, which reshapes the skyrmion texture through propagation.

These two examples highlight that distinct coherence structures can drive qualitatively different topological transitions, thereby establishing spatial coherence engineering as a versatile route for controlling skyrmion states of light. The corresponding numerical results for the two types of partially coherent skyrmion beams with non-Gaussian correlation agree well with the experimental observations (Supplementary Fig.~5). The coherence-induced transitions reported here are also applicable to other partially coherent skyrmion beams; for example, analogous behavior is obtained for a beam composed of partially coherent LG$_{0}^0$ and LG$_{0}^2$ modes (Supplementary Fig.~6).

\section{Discussion}
In this work we have investigated how partial spatial coherence reshapes, stabilizes, and controls optical skyrmions encoded in polarization textures during propagation in both free space and atmospheric turbulence. A central difficulty in extending skyrmion topology from ideal coherent beams to realistic stochastic fields is the emergence of propagation-induced, spatially inhomogeneous depolarization, which causes the conventional Stokes-vector mapping to collapse into the interior of the Poincar\'e sphere and obscures a well-defined topological characterization. To address this, we introduced a reshaped parameter-space representation based on a local renormalization of the Stokes vector, enabling a consistent definition of partially coherent skyrmion fields and their skyrmion number in the presence of nonuniform polarization degree. With this framework, we identified a distinct coherence-enabled regime in which the skyrmion number remains stable during free-space propagation, and we further demonstrated that reducing spatial coherence can markedly enhance the robustness of skyrmionic topology in atmospheric turbulence. Importantly, this stabilization persists even in the strong-fluctuation regime (scintillation index exceeding unity), where fully coherent skyrmions exhibit pronounced degradation and realization-to-realization instability. These findings suggest a viable route toward real-time, topology-encoded free-space photonic links, where information is carried by skyrmion number rather than by fragile field details.

Beyond robustness, we show that spatial coherence engineering constitutes an independent and powerful control degree of freedom for topological transitions. By tailoring non-Gaussian coherence landscapes with phase discontinuities, we realize propagation-induced topological transitions, including skyrmion-to-skyrmionium conversion and controlled splitting into skyrmion lattices. Notably, these transformations require no additional polarization components along the propagation path; instead, the desired topology is programmed at the source through the coherence structure and emerges spontaneously through propagation. This coherence-programmable approach opens opportunities for reconfigurable topological beam shaping and multiplexing, with potential relevance to robust structured-light communications, topological metrology, and sensing in complex environments.

More broadly, our study highlights an overarching message: partial coherence, ubiquitous in natural and practical optical fields, is not merely an imperfection to be mitigated but a resource that can protect and even engineer topological states of light. Extending skyrmions from pristine lasers to stochastic photonic fields therefore enriches topological photonics with the tools and insights of statistical optics, and points to coherence-engineered topological textures as a practical platform for resilient photonic functionalities in the real world.

\section*{Methods}

\subsection{Optical skyrmions in the Stokes-vector texture}
The normalized Stokes vector of an optical beam is defined as
\begin{equation}
\label{eqSS}
\mathbf{S}(\mathbf{r})=\frac{1}{s_0(\mathbf{r})}\big[s_1(\mathbf{r}),\,s_2(\mathbf{r}),\,s_3(\mathbf{r})\big],
\end{equation}
where $s_j(\mathbf{r})$ ($j=0,1,2,3$) are the Stokes parameters evaluated at the transverse position $\mathbf r$. In terms of the local polarization matrix $\boldsymbol{\Phi}(\mathbf{r})= \mathbf{E}^\ast(\mathbf{r}) \mathbf{E}^\mathrm{T}(\mathbf{r})$, the Stokes parameters can be written as
\begin{equation}
s_j(\mathbf{r})=\mathrm{Tr}\!\left[\boldsymbol{\sigma}_j\,\boldsymbol{\Phi}(\mathbf{r})\right],
\end{equation}
where $\boldsymbol{\sigma}_0$ is the $2\times2$ identity matrix and $\boldsymbol{\sigma}_{1,2,3}$ are the Pauli matrices.

The local degree of polarization, which quantifies the fraction of fully polarized component in the total field at $\mathbf r$, is given by
\begin{equation}
P(\mathbf{r})=\frac{\sqrt{s_1^2(\mathbf{r})+s_2^2(\mathbf{r})+s_3^2(\mathbf{r})}}{s_0(\mathbf{r})}.
\end{equation}

Following Ref.~\cite{Shen2024NP}, the Stokes texture of an optical skyrmion can be conveniently parameterized as
\begin{equation}
\label{eqS}
\mathbf{S}(\mathbf{r})=
\big[\cos\alpha(\theta)\sin\beta(r),\;
\sin\alpha(\theta)\sin\beta(r),\;
\cos\beta(r)\big],
\end{equation}
where $\alpha(\theta)$ describes the azimuthal winding of the transverse Stokes components and $\beta(r)$ determines the radial evolution from the beam center to the boundary of the skyrmion domain.
With this representation, the skyrmion number can be factorized as $N_{\mathrm{sk}}=q\,m$, where
\begin{equation}
q=\frac{1}{2}\Big[\cos\beta(r)\Big]_{r=0}^{r=r_\sigma},
\qquad
m=\frac{1}{2\pi}\Big[\alpha(\theta)\Big]_{\theta=0}^{2\pi}.
\end{equation}
Here $q$ is the polarity, determined by whether $S_3$ points down (up) at the center and up (down) at the boundary (corresponding to $q=+1$ or $q=-1$), and $m$ is the vorticity associated with the azimuthal winding of the transverse Stokes field $\mathbf{S}_{\perp}=[S_1,S_2]$.
By tailoring the polarization basis, spatial modes, and the relative phase $\gamma$, a variety of Stokes-vector topological textures can be constructed, including Bloch- and N\'eel-type skyrmions, anti-skyrmions, merons, and bimerons \cite{Shen2024NP}.

\subsection{Experimental set-up and description}
The schematic of the experimental system for generating partially coherent skyrmion beams and measuring their Stokes parameters is shown in Fig.~\ref{figure7}. A linearly polarized continuous-wave laser ($\lambda=532~\mathrm{nm}$) is expanded by a beam expander (BE, $20\times$) and then split by a $50:50$ beam splitter (BS$_1$). One output arm illuminates a phase-only spatial light modulator (SLM; $1920\times1080$ pixels, pixel pitch $8~\mu\mathrm{m}$, refresh rate $56~\mathrm{Hz}$). The SLM display is divided into two halves, each playing a time sequence of computer-generated holograms (CGHs) to generate random field realizations. Time averaging over these random realizations yields partially coherent Laguerre--Gaussian modes: LG$_0^0$ from the left half and LG$_0^l$ from the right half of the SLM screen. Details of the CGH synthesis are given in Sec.~C.

After reflection from the SLM, the field enters a common-path $4f$ system formed by lenses L$_1$ and L$_2$ (identical focal lengths). A V-shaped aperture (VA) placed at the Fourier plane between L$_1$ and L$_2$ suppresses the zero order and unwanted diffraction orders, transmitting only the two desired first-order beams. The two beams then pass through independent polarization-control elements (HWP$_1$ and HWP$_2$), which convert their linear polarizations into right- and left-circular polarizations, respectively. A Ronchi grating (RG) placed after the $4f$ system spatially overlaps the two beams, producing a single vector partially coherent skyrmion beam at the RG output.

In our experiment, $1000$ random modes are used to synthesize the partially coherent skyrmion beam, and these modes are displayed cyclically on the SLM. The spatial distributions of these random modes control both the coherence and topological structures of the synthesized beam. We note that this SLM-based implementation is intended as a proof-of-principle demonstration of coherence-engineered skyrmion generation, rather than as a real-time source for practical turbulence communication experiments. In a real-time turbulence scenario, the source fluctuation time should be much shorter than the characteristic turbulence evolution time, which is typically on the scale of tens of milliseconds \cite{WangH2025}. Therefore, the SLM used in our experiment is too slow for this specific real-time turbulence condition. A practical implementation could instead use a much faster spatial modulator, such as a digital micromirror device (DMD), which can operate at kHz to tens-of-kHz rates \cite{WangH2025, WangY2024}. For example, with $1000$ random patterns and a refresh rate in the tens-of-kHz range, the synthesis time can be reduced to the millisecond scale, well below the characteristic evolution time of atmospheric turbulence. Thus, while the present SLM experiment demonstrates the principle of coherence-engineered skyrmion generation, faster modulators would be required for real-time applications in dynamic turbulent channels.

To characterize the skyrmion source plane (immediately after the RG), an imaging lens L$_3$ relays the source plane onto a CCD camera with unit magnification. A polarization analyzer, consisting of a quarter-wave plate followed by a linear polarizer, is placed in front of the CCD to perform conventional Stokes polarimetry (SP) and retrieve $s_0$--$s_3$, from which the Stokes texture and skyrmion number are computed (see Sec.~D). To measure the propagation dynamics, the analyzer and CCD are mounted on a motorized translation stage and scanned along $z$, allowing acquisition of the Stokes parameters at different propagation distances.

\subsection{Hologram design}
To generate two partially coherent orthogonally polarized modes simultaneously, the SLM is divided into left and right halves that encode, respectively, the target deterministic fields $E_\mathrm{left}(\mathbf r)$ (LG$_0^0$) and $E_\mathrm{right}(\mathbf r)$ (LG$_0^l$). To realize the partially coherent vector beam consistent with Eq.~(\ref{eqCSD}), we impose the same (time-varying) random complex screen $T_n(\mathbf r)$ on both halves,
\begin{align}
\label{eqEn}
E_{\mathrm{left},n}(\mathbf r) &= E_\mathrm{left}(\mathbf r)\,T_n(\mathbf r), \nonumber \\
E_{\mathrm{right},n}(\mathbf r) &= E_\mathrm{right}(\mathbf r)\,T_n(\mathbf r),
\end{align}
where $n=1,2,\ldots$, so that the two polarization components remain fully correlated within each realization. The SLM output remains the same linear polarization for two modes; the conversion to left- and right-circular polarizations is implemented downstream by HWP$_1$ and HWP$_2$. The random screens $T_n(\mathbf r)$ are synthesized assuming Gaussian statistics. Specifically, we generate complex Gaussian white noise $R_n(\mathbf f)$ (zero mean and unit variance) in the spatial-frequency domain and filter it by the square root of the desired power spectrum $p(\mathbf f)$, yielding
\begin{equation}
T_n(\mathbf r)=\mathcal{F}^{-1}\!\left[R_n(\mathbf f)\sqrt{p(\mathbf f)}\right],
\end{equation}
where $\mathcal{F}^{-1}$ denotes the inverse Fourier transform. The correlation of the screens satisfies $\mu(\mathbf r_1,\mathbf r_2)\propto \langle T_n^\ast(\mathbf r_1)T_n(\mathbf r_2)\rangle$, and thus determines the prescribed degree of coherence. The spatial coherence structure thus is controlled by the power spectrum $p(\mathbf f)$. For each realization $n$, the complex fields $E_{\mathrm{left},n}(\mathbf r)$ and $E_{\mathrm{right},n}(\mathbf r)$ are encoded into a phase-only CGH following the method in Ref.~\cite{Guzman2017}. The resulting CGH sequence is displayed cyclically on the SLM to realize a temporally varying ensemble of vector-field realizations.

\subsection{Stokes polarimetry}
The Stokes texture, skyrmion density, and skyrmion number are reconstructed from experimentally measured Stokes parameters. We perform conventional Stokes polarimetry (SP) using a quarter-wave plate (QWP) and a linear polarizer placed before the CCD. By setting the QWP and polarizer at appropriate angles, we record the time-averaged intensity profiles in six polarization channels: horizontal ($I_\mathrm{H}$), vertical ($I_\mathrm{V}$), diagonal ($I_\mathrm{D}$), anti-diagonal ($I_\mathrm{A}$), right-circular ($I_\mathrm{R}$), and left-circular ($I_\mathrm{L}$). The Stokes parameters are then obtained as
\begin{align}
\label{eqeEStokes}
s_0(\mathbf r) = I_\mathrm{R}(\mathbf r)+I_\mathrm{L}(\mathbf r), ~
s_1(\mathbf r) &= I_\mathrm{H}(\mathbf r)-I_\mathrm{V}(\mathbf r),\nonumber \\
s_2(\mathbf r) = I_\mathrm{D}(\mathbf r)-I_\mathrm{A}(\mathbf r), ~
s_3(\mathbf r) &= I_\mathrm{R}(\mathbf r)-I_\mathrm{L}(\mathbf r).
\end{align}
Representative measurements of the six polarization-resolved intensity profiles for a partially coherent skyrmion beam composed of LG$_0^0$ and LG$_0^1$ spatial modes are shown in the inset of Fig.~\ref{figure7}.

\section*{Data Availability}
The data supporting the findings of this study are available within the paper and its Supplementary Information. The corresponding authors can also provide the data upon request.

\section*{Code Availability}
The code used to produce the results are available from the corresponding authors upon request.

\section*{Acknowledgements}
The authors thanks the reviewers for their constructive comments and valuable suggestions.

\section*{Funding}
Y.J.C. discloses support for the research of this work from the National Natural Science Foundation of China (grants 12192254, 12534014, W2441005), the National Key Research and Development Program of China (2022YFA1404800), the Natural Science Foundation of Shandong Province (ZR2025ZD21), and the Key Research and Development Program of Shandong Province (2024JMRH0105). Y.L. discloses support for the research of this work from the National Natural Science Foundation of China (12404348). Y.H.C. discloses support for the research of this work from the National Natural Science Foundation of China (12274310). F.W. discloses support for the research of this work from the National Natural Science Foundation of China (12274311). Y.S. discloses support for the research of this work from the Singapore Ministry of Education (MOE) AcRF Tier 1 (RG157/23 \& RT11/23), the Singapore Agency for Science, Technology and Research (A*STAR) (M24N7c0080 \& R25J4IR110), and the Nanyang Assistant Professorship Start Up grant. The other authors (S.C, Z.G, and K.Z) declare no relevant funding.

\section*{Author Contributions}
F.W, Y.S, Y.H.C and Y.L conceived the idea; The design, theoretical analysis and numerical simulation were carried out by Y.L, and F.W, which was verified in experiment by Y.L; Y.L collected the data, and Y.L, Z.G, S.C, K.Z, Y.H.C, Y.J.C, Y.S and F.W carried out the data analysis; Y.L, Y.H.C, and F.W contributed to writing--original draft and all authors contributed to the  writing--review and editing; The entire research was supervised by Y.H.C, Y.S, Y.J.C, and F.W; All authors have read and agreed to the published version of the manuscript.

\section*{Competing interests}
The authors declare no competing interests.\\

\section*{Figure Legends/Captions}

\begin{figure*}[h]
\centering
\includegraphics[width=0.86 \linewidth]{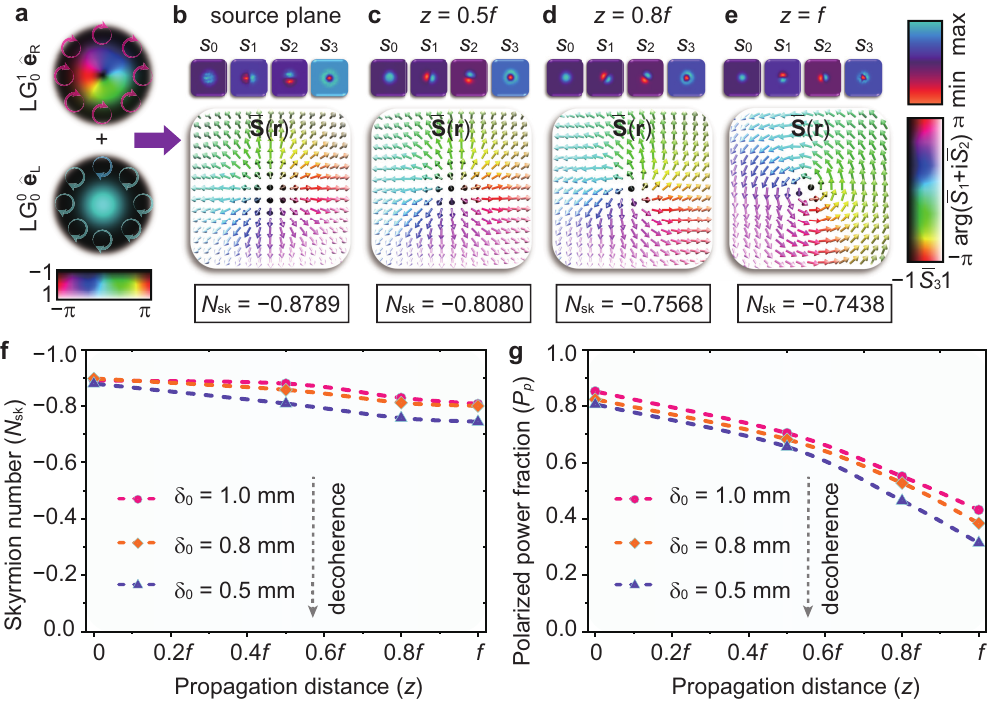}
\caption{\textbf{Propagation dynamics of a partially coherent skyrmion beam with a Gaussian degree of spatial coherence}. (a) The input polarization texture is synthesized by coherently superposing two orthogonally polarized modes: a right-handed $\mathrm{LG}^{1}_{0}$ and a left-handed $\mathrm{LG}^{0}_{0}$. (b)--(e) Measured transverse distributions of the unnormalized Stokes parameters $s_0$--$s_3$ (top rows) and the corresponding normalized Stokes texture $\overline{\mathbf{S}}(\mathbf{r})$ (bottom rows) at the source plane and at selected propagation distances $z$ during focusing by a thin lens with focal length $f$. The skyrmion number $N_\mathrm{sk}$ extracted from each texture is indicated below. Evolution of (f) the skyrmion number $N_\mathrm{sk}$ and (g) the polarized power fraction $P_p$, defined as the ratio of the fully polarized power to the total power, as a function of propagation distance for different coherence widths $\delta_0$.}
\label{figure2}
\end{figure*}

\begin{figure*}[h!]
\centering
\includegraphics[width= 0.88 \linewidth]{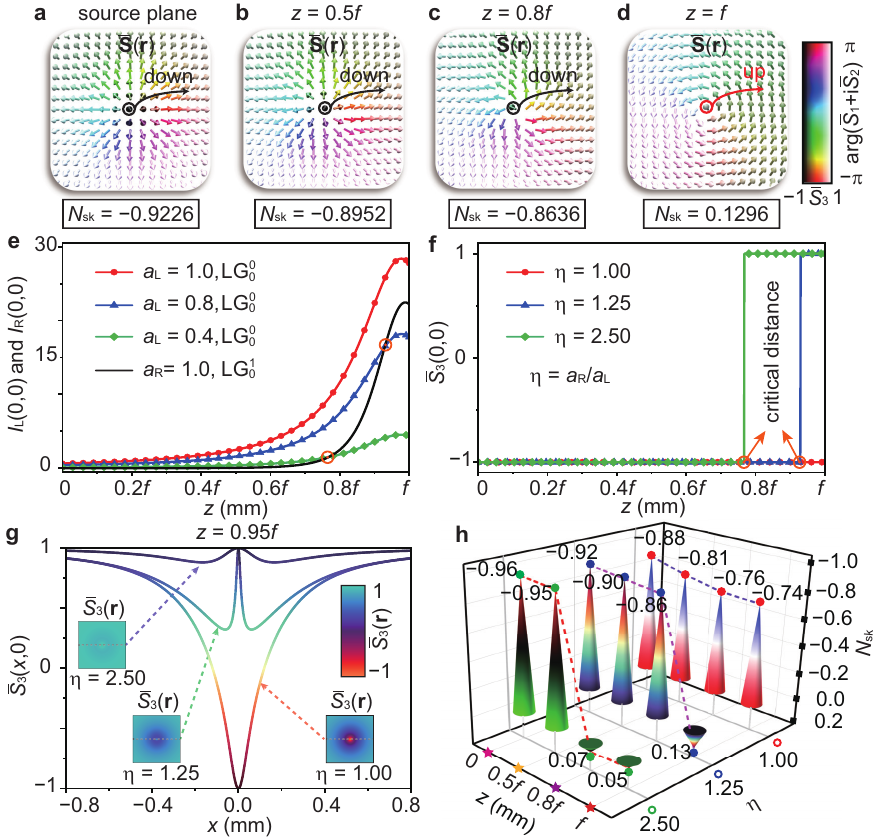}
\caption{\textbf{Amplitude-ratio-dependent robustness and breakdown of partially coherent skyrmion textures}. (a)--(d) Evolution of the normalized Stokes texture $\overline{S}(\mathbf{r})$ at the source plane and at selected propagation distances during focusing (Gaussian spatial coherence), shown here for $\eta=1.25$. The indicated labels ``down'' and ``up'' mark the sign of the central $\overline{S}_3$ component, and the skyrmion number $N_\mathrm{sk}$ extracted from each texture is given below. (e) Evolution of the on-axis intensities of the partially coherent LG$_0^0$ and LG$_0^1$ components during propagation for three different amplitude ratios. (f) Evolution of the central Stokes component $\overline{S}_3(0,0)$ during propagation. The critical propagation distances at which the on-axis intensity of the partially coherent LG$_0^1$ component exceeds that of the LG$_0^0$ component, leading to the reversal of $\overline{S}_3(0,0)$, are highlighted by circles. (g) Cross-sectional profiles of $\overline{S}_3(x,0)$ at $z=0.95f$ for representative amplitude ratios $\eta=1.00$, $1.25$, and $2.50$, illustrating the on-axis crossover that leads to a central $\overline{S}_3$ sign reversal when $\eta>1$ (insets: corresponding $\overline{S}_3(\mathbf{r})$ maps). (h) Measured skyrmion number $N_\mathrm{sk}$ as a function of propagation distance and amplitude ratio $\eta$, showing accelerated degradation of topology for $\eta>1$ and robust preservation for $\eta\leq 1$.}
\label{figure3}
\end{figure*}

\begin{figure*}[h!]
\centering
\includegraphics[width= 0.9 \linewidth]{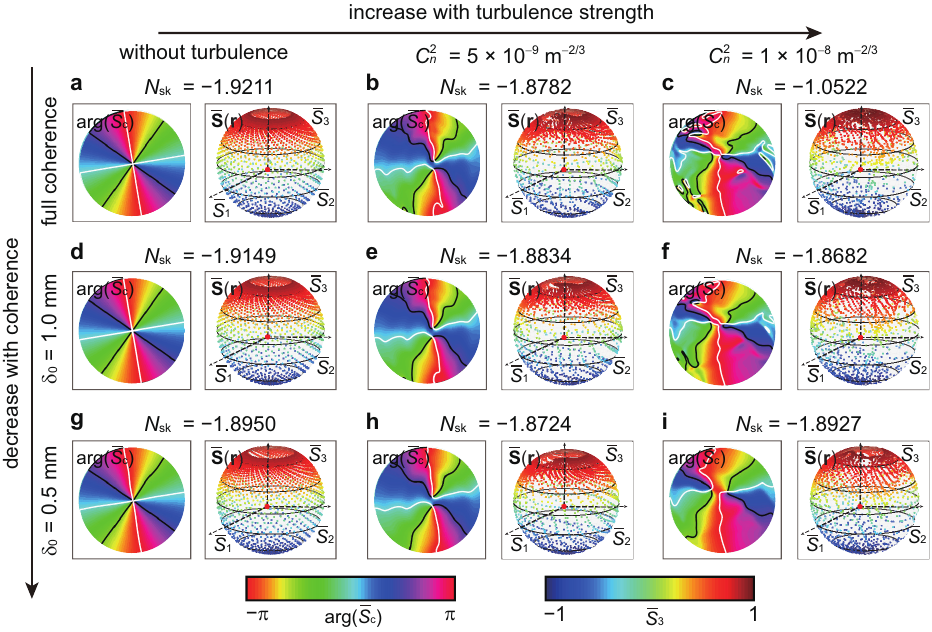}
\caption{\textbf{Turbulence-induced singularity proliferation and coherence-enabled stabilization of skyrmion topology}. Columns correspond to increasing turbulence strength (from left to right): without turbulence, $C_n^2=5\times10^{-9}~\mathrm{m^{-2/3}}$, and $C_n^2=1\times10^{-8}~\mathrm{m^{-2/3}}$. Rows correspond to decreasing spatial coherence (from top to bottom): fully coherent input, partially coherent input with $\delta_0=1.0~\mathrm{mm}$, and $\delta_0=0.5~\mathrm{mm}$. For each case, each panel show the phase of the complex Stokes field $\arg(\overline{S}_\mathrm{c})$ with $\overline{S}_\mathrm{c}=\overline{S}_1+i\overline{S}_2$ (left) and the corresponding mapping of the renormalized Stokes vectors $\overline{\mathbf S}(\mathbf r)$ on the reshaped parameter sphere (right). The extracted skyrmion number $N_\mathrm{sk}$ is indicated above each panel. With increasing turbulence, diffraction generates additional phase singularities in $\overline{S}_\mathrm{c}$ and leads to an increasingly irregular parameter-sphere mapping for the fully coherent beam (a--c). Reducing the spatial coherence suppresses the singularity proliferation and restores a more uniform mapping, thereby stabilizing the skyrmion topology (d--i).}
\label{figure4}
\end{figure*}

\begin{figure*}[h!]
\centering
\includegraphics[width= 0.94 \linewidth]{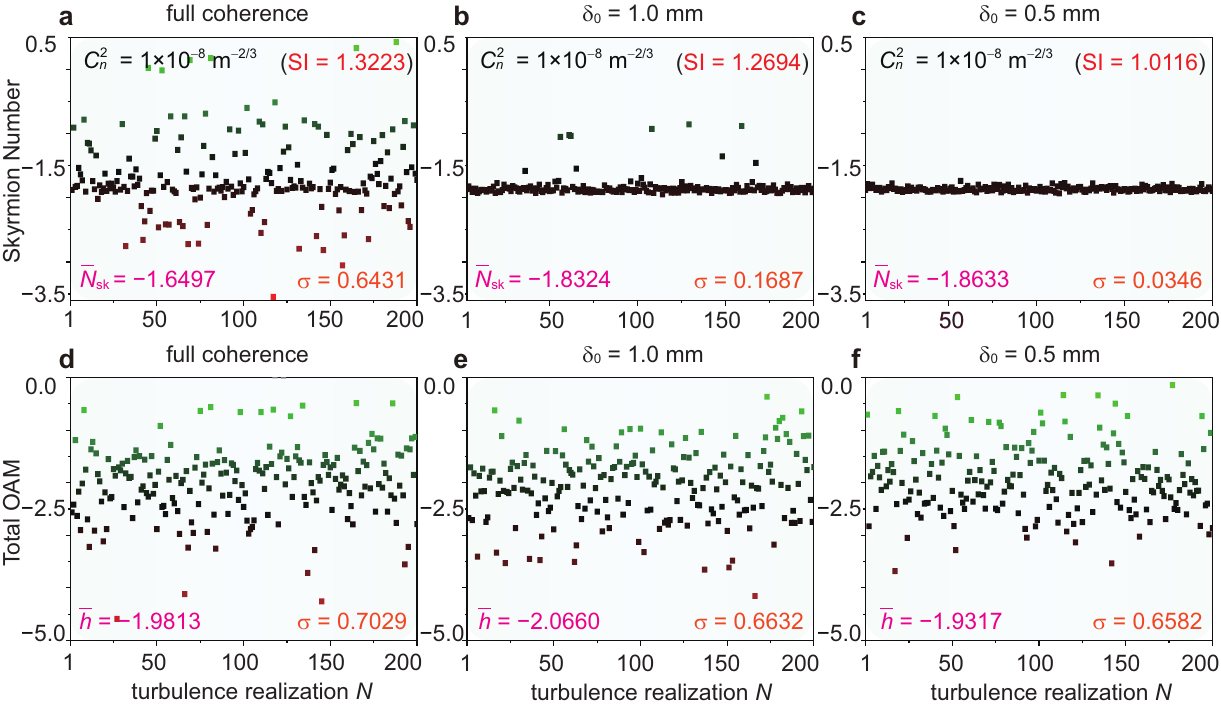}
\caption{\textbf{Coherence-enhanced stability of the skyrmion number in strong atmospheric turbulence and comparison with total OAM}. (a--c) Skyrmion number $N_\mathrm{sk}$ evaluated at the receiver plane for $200$ independent turbulence realizations for (a) a fully coherent beam, (b) a partially coherent beam with $\delta_0=1.0~\mathrm{mm}$, and (c) a partially coherent beam with $\delta_0=0.5~\mathrm{mm}$. The mean value $\overline{N}_\mathrm{sk}$ and standard deviation $\sigma$ are annotated in each panel. (d--f) Corresponding statistics of the total orbital angular momentum (OAM) per photon for the same turbulence realizations and coherence widths; the mean $\overline{h}$ and standard deviation $\sigma$ are shown. All panels correspond to strong turbulence with $C_n^2=1\times10^{-8}~\mathrm{m^{-2/3}}$. The scintillation indices (SI) of the received intensity for the three different coherence widths are highlighted in the figure.}
\label{figure5}
\end{figure*}

\begin{figure*}[h!]
\centering
\includegraphics[width= 0.88 \linewidth]{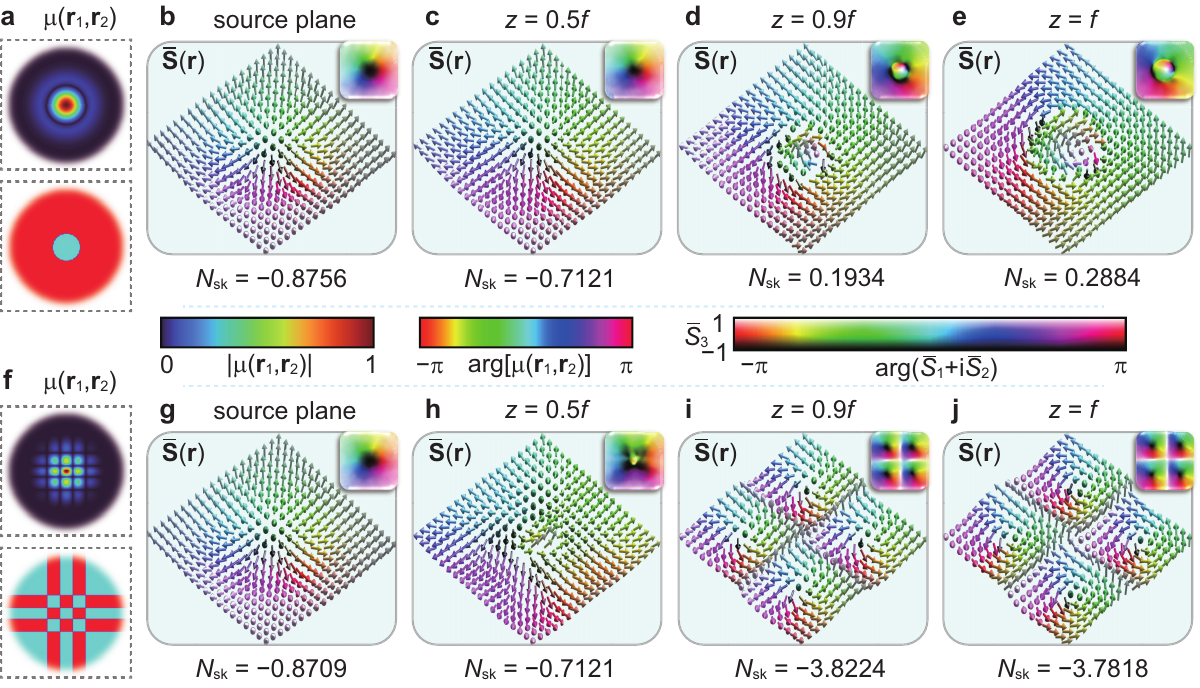}
\caption{\textbf{Coherence-driven topological transitions of partially coherent optical skyrmions enabled by non-Gaussian spatial correlations}. (a) Amplitude $|\mu(\mathbf r_1,\mathbf r_2)|$ and phase $\arg[\mu(\mathbf r_1,\mathbf r_2)]$ of a Laguerre--Gaussian (LG) degree of coherence. (b--e) Propagation-induced skyrmion-to-skyrmionium transition under the LG coherence: shown are the normalized Stokes textures $\overline{\mathbf S}(\mathbf r)$ at the source plane and at selected propagation distances during focusing ($z=0$, $0.5f$, $0.9f$, and $f$). The extracted skyrmion number $N_\mathrm{sk}$ is indicated below each texture. Insets encode $\arg(\overline {S}_1+i\overline {S}_2)$ and $\overline {S}_3$ as indicated by the color bars. (f) Amplitude and phase of a Hermite--Gaussian (HG) degree of coherence. (g--j) Coherence-enabled skyrmion splitting and multiplication under the HG coherence: shown are the normalized Stokes textures $\overline{S}(\mathbf r)$ at $z=0$, $0.5f$, $0.9f$, and $f$, where the initial skyrmion texture evolves into four spatially separated skyrmionic parts, resulting in an increased magnitude of the total skyrmion number.}
\label{figure6}
\end{figure*}

\begin{figure*}[h!]
\centering
\includegraphics[width= 0.87 \linewidth]{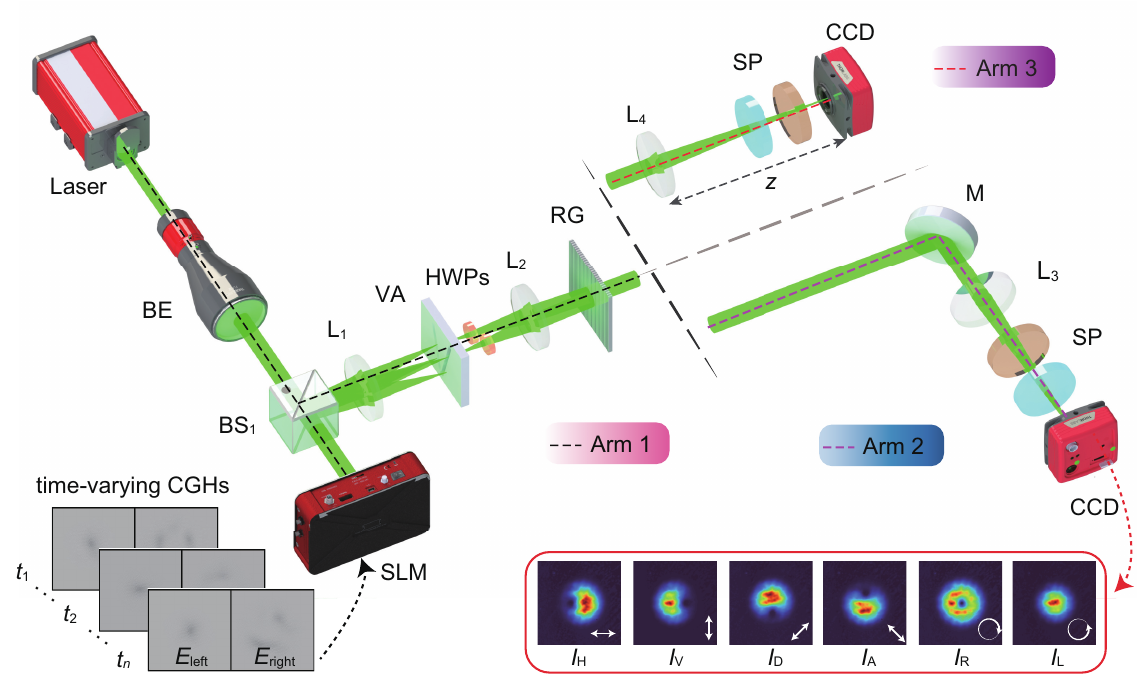}
\caption{\textbf{Experimental set-up for generating partially coherent skyrmion beams and measuring their Stokes parameters}. A linearly polarized laser is expanded by a beam expander (BE) and split by a $50{:}50$ beam splitter (BS$_1$). One arm illuminates a spatial light modulator (SLM) displaying time-varying computer-generated holograms (CGHs) on two halves of the screen to generate two partially coherent scalar fields $E_{\mathrm{left}}$ and $E_{\mathrm{right}}$ sharing the same random-screen realization. The modulated beams are processed by a common-path $4f$ system (L$_1$--L$_2$) with a V-shaped aperture (VA) to select the desired diffraction orders, and their polarizations are converted by half-wave plates (HWPs). A Ronchi grating (RG) recombines the two components into a single vector partially coherent skyrmion source. Stokes polarimetry (SP; quarter-wave plate + linear polarizer) together with a CCD camera is used to record six polarization-resolved intensity images ($I_\mathrm{H}$, $I_\mathrm{V}$, $I_\mathrm{D}$, $I_\mathrm{A}$, $I_\mathrm{R}$, $I_\mathrm{L}$) for reconstructing the Stokes parameters. Lens L$_3$ images the source plane onto the CCD (Arm~2), and a second measurement arm (Arm~3, L$_4$) translates along $z$ to acquire Stokes parameters at different propagation distances.}
\label{figure7}
\end{figure*}

\end{document}